\documentclass[12pt]{article}

\input epsf
\usepackage{amssymb}
\usepackage[dvips]{graphicx}
\usepackage{amscd}
\usepackage[spanish]{babel}

\long\def\symbolfootnote[#1]#2{\begingroup%
\def\thefootnote{\fnsymbol{footnote}}\footnote[#1]{#2}\endgroup}

\def\dj{\hbox{d\kern-0.347em \vrule width 0.3em height 1.252ex depth
-1.21ex \kern 0.051em}}

\def\ee{{\rm e}\,}

\def\a{{\'a}}
\def\e{{\'e}}
\def\ii{{\'{\i}}}
\def\oo{{\'o}}
\def\u{{\'u}}
\def\nn{{\~n}}
\def\A{{\'A}}
\def\E{{\'E}}
\def\I{{\'{\I}}}
\def\be{\begin{equation}}
\def\ee{\end{equation}}

\begin{document}

\begin{flushright}
CERN-PH-TH/034-2005 \\
{\tt physics/0503150}
\end{flushright}

\vspace*{1cm}

\centerline{\bf \Large Einstein y la Geometr\ii a}

\vspace*{1cm}

\centerline{Luis \A lvarez-Gaum\e$^{\,\rm a,}$\symbolfootnote[1]{
{\tt Luis.Alvarez-Gaume@cern.ch}} y Miguel A. 
V\a zquez-Mozo$^{\,\rm b,}$\symbolfootnote[3]{{\tt 
Miguel.Vazquez-Mozo@cern.ch}}}
\vspace*{0.4cm}
\begin{quote}
$^{\,\rm a\,}$ {\sl Physics Department, Theory Division CERN, CH-1211 Geneva 23,
Switzerland} 
\\
%\noindent
$^{\,\rm b\,}$ {\sl F\ii sica Te\oo rica,  
Universidad de Salamanca, Plaza de la Merced s/n, E-37008 Salamanca,
Spain}
\end{quote}

\vspace*{1cm}

\section{Introducci\oo n}

La Geometr\ii a fue la primera rama de las Matem\a ticas en recibir
una formulaci\oo n sistem\a tica hace m\a s de dos mil a\nn os.  En
``Los Elementos'' de Euclides, escrito alrededor del a\nn o 300 antes
de nuestra era y unos de los libros m\a s influyentes en la historia
de la Humanidad, se obtienen los principales teoremas de la geometr\ii
a plana a partir de cinco postulados. Aunque el trabajo de Euclides
domin\oo\ la Geometr\ii a durante siglos, fueron numerosos los matem\a
ticos desde Ptolomeo a Legendre que intentaron demostrar que el quinto
y \u ltimo de los postulados introducidos por Euclides\footnote{El
quinto postulado de Euclides es equivalente al llamado postulado de
Playfair que afirma que en el plano, dado una l\ii nea recta y un
punto externo a ella existe una sola recta paralela a la primera y que
pase por dicho punto.}  pod\ii a o bien deducirse de los otros cuatro
postulados euclidianos o reducirse a una proposici\oo n m\a s
``simple''.  S\oo lo en el siglo XIX los matem\a ticos Carl Friedrich
Gauss, Nikolai Lobachevski y J\a nos Bolyai se dieron cuenta,
independientemente, de que el famoso quinto postulado no s\oo lo era
independiente de los cuatro restantes sino que pod\ii a relajarse para
dar lugar a geometr\ii as no euclidianas perfectamente
autoconsistentes.  Estas nuevas geometr\ii as fueron generalizadas
poco despu\e s por Bernhard Riemann a un n\u mero arbitrario de
dimensiones, inaugurando as\ii\ toda una nueva rama de las Matem\a
ticas.

El objetivo de este art\ii culo es mostrar al lector como a trav\e s
del trabajo de Einstein las geometr\ii as no euclidianas, que en sus
or\ii genes pod\ii an parecer exentas de toda relevancia f\ii sica, 
se convirtieron en uno de los cimientos sobre los que descansa nuestro 
conocimiento del Universo y la gravitaci\oo n. Para ello, 
en lugar de presentar disquisiciones eruditas
sobre los desarrollos que la teor\ii a de la relatividad ha generado e
inspirado en el campo de la Geometr\ii a del siglo XX (que han sido
muchos), nos gustar\ii a discutir de la forma m\a s accesible posible
cuales fueron las ideas con las que Einstein convirti\oo\ a la geometr\ii a
no euclidiana en un lenguaje natural para la F\ii sica.

\section{La geometr\ii a del espacio-tiempo}

La geometr\ii a ha influido muy poderosamente el desarrollo de las
Ciencias F\ii sicas, no solamente en la antig\"uedad sino durante todo
el proceso que culmin\oo{} con la creaci\oo n de la F\ii sica como
ciencia experimental en el siglo XVII.  Esta influencia se
manifest\oo{} fundamentalmente en el uso de la geometr\ii a como
herramienta de an\a lisis de los procesos f\ii sicos, una de las
razones por las cuales el lector contemporaneo encuentra dificil la
lectura de libros como los {\it Principia} newtonianos. Son m\a s
raros, sin embargo, los intentos de ``geometrizar'' la F\ii sica, es
decir, reducir las leyes f\ii sicas a propiedades geom\e tricas del
espacio.

Quiz\a s el m\a s c\e lebre intento en este sentido fue el plan de
Johannes Kepler de desentra\nn ar el ``Misterio del Universo''
explicando los tama\nn os de las \oo rbitas planetarias en t\e rminos
puramente geom\e tricos. Aunque hoy sabemos que los tama\nn os de las
\oo rbitas en el Sistema Solar tienen los valores que
tienen por razones puramente ``hist\oo ricas'', el programa kepleriano dio
como resultado las tres leyes del movimiento planetario, uno de los
pies del gigante a cuyos hombros Newton fue capaz de atisbar m\a s
lejos que nadie antes que \e l.

Tuvieron que pasar no obstante 300 a\nn os desde el intento de Kepler
para que la geometrizaci\oo n de la F\ii sica viniera de la mano de la
teor\ii a de la relatividad einsteiniana.  A pesar de la fascinaci\oo n
infantil de Einstein con la geometr\ii a \cite{NA} no fue \e l quien
dio el primer paso. En su celeb\e rrimo art\ii culo de junio de 1905
\cite{1905} la relatividad es presentada de manera f\ii sica a trav\e
s de relojes y reglas. Fue el matem\a tico suizo Hermann Minkowski
qui\e n en 1908 se dio cuenta de que las transformaciones entre los
diferentes sistemas de referencia inerciales se pod\ii an entender
geom\e tricamente como ciertos cambios de coordenadas en el
``espacio-tiempo'', un hiperespacio de cuatro dimensiones en
las tres de ellas se identifican con las dimensiones espaciales
habituales mientras que la cuarta corresponde al tiempo.

\begin{figure}[ht]
\centerline{ \epsfxsize=3.0truein \epsfbox{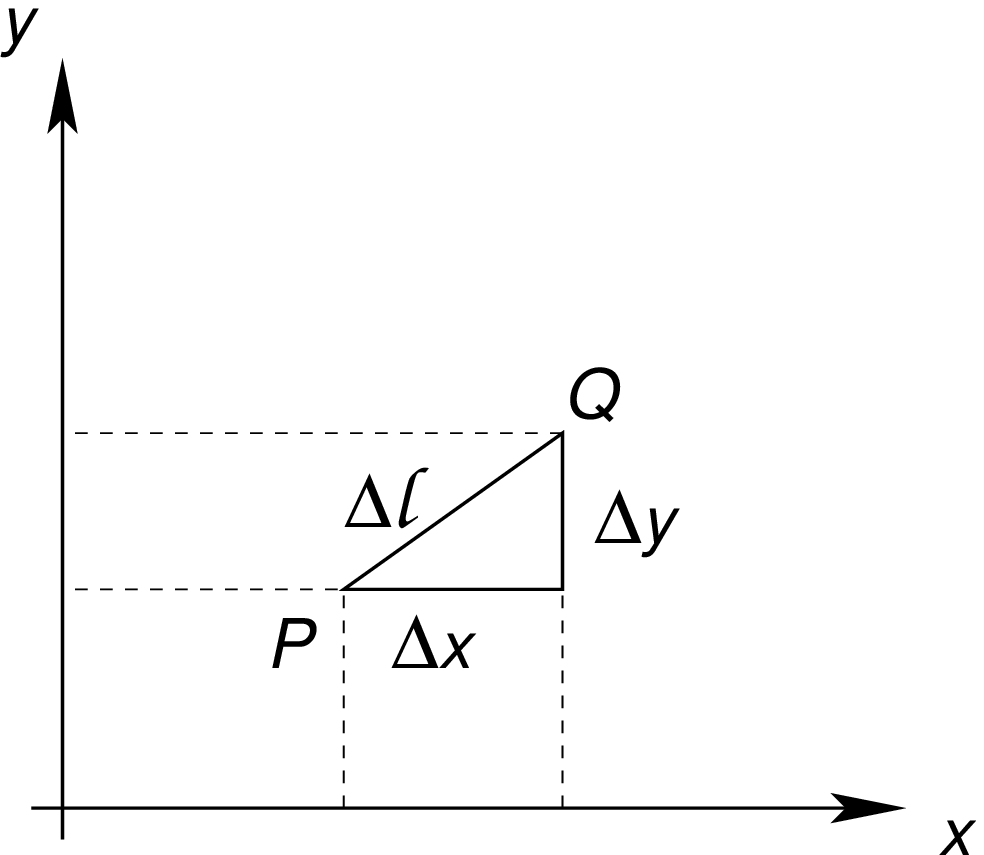}}
\caption{Ilustraci\oo n del teorema de Pit\a goras.}
\label{1}
\end{figure}

La clave de la geometr\ii a no euclidiana es el
hecho de que en un entorno suficientemente peque\nn o de cada punto la
geometr\ii a es muy aproximadamente euclidiana. En particular las
distancias pueden calcularse directamente utilizando el teorema de Pit\a
goras. Entre dos puntos cercanos $P$ y $Q$ cuya diferencia de
coordenadas es $\Delta x$ y $\Delta y$, la distancia viene dada
por (ver figura \ref{1})
\begin{equation}\label{pitagoras}
\Delta \ell^2 = \Delta x^2 + \Delta y^2.
\end{equation}
De esta forma la geometr\ii a local en el plano no eucl\ii deo
siempre est\a\ dada por la ecuaci\oo n anterior. Esta idea condujo a la
fecunda noci\oo n de variedad riemanniana, cuyo desarrollo ha tenido un
impacto notable en la historia de las matem\a ticas.

Cuando hablamos aqu\ii\ de geometr\ii as no euclidianas nos estamos
refiriendo en todo momento al espacio. El tiempo en la f\ii sica cl\a
sica, as\ii\ como en matem\a ticas, es siempre un espectador inmutable
que juega el papel de un par\a metro externo.  Esto es particularmente
obvio cuando consideramos el principio de la relatividad formulado por
Galileo y que jug\oo\ un papel fundamental tanto en la mec\a nica de
Newton como en su extensi\oo n llevada a cabo por Einstein.

El principio de relatividad galileano supone la existencia de un
conjunto infinito de sistemas de referencia inerciales (esto es,
sistemas de referencia en los que se satisface la ley de inercia) y
que necesariamente se mueven unos con respecto a los otros con
velocidad constante.  Este principio afirma que las leyes de la mec\a
nica toman la misma forma en todos los sistemas inerciales.

El principio de la relatividad de Galileo se puede entender tambi\e n
como un principio de invariancia. El paso de la descripci\oo n en un
sistema de referencia inercial a otro se realiza a trav\e s de ciertas
transformaciones que relacionan las coordenadas y el tiempo en ambos
sistemas, las llamadas transformaciones de Galileo. El principio de
relatividad galileano se enuncia entonces diciendo que las leyes de la
mec\a nica son invariantes en forma frente a dichas transformaciones.
En particular, en la relatividad galileana el tiempo no cambia al
pasar de un sistema de referencia inercial a otro m\a s all\a\ del
cambio producido por una diferente elecci\oo n en el origen de tiempos:
los relojes de dos observadores inerciales no tienen por qu\e\ marcar
la misma hora, pero s\ii\ medir\a n el mismo intervalo temporal
transcurrido entre dos sucesos.

Este es quiz\a s el punto m\a s importante donde la relatividad
einsteiniana se separa de y extiende a la galileana. El principio de
la relatividad de Einstein se basa en dos postulados fundamentales:
\begin{itemize}
\item
{\em Todas} las leyes de la f\ii sica toman la misma forma en cualquier
sistema de referencia inercial.
\item
La velocidad de la luz es independiente de la velocidad relativa del 
observador y la fuente.
\end{itemize}
Es decir, no s\oo lo las leyes de la mec\a nica sino todas la leyes de
la f\ii sica, y en particular las del electromagnetismo, tienen que
ser invariantes bajo las transformaciones que relacionan coordenadas y
tiempo en dos sistemas de referencia inerciales. Adem\a s, la
velocidad de la luz emitida por una fuente que medir\ii an diferentes
observadores inerciales ha de ser siempre la misma (unos 300.000 km
s$^{-1}$).

Si intentamos obtener las transformaciones de coordenadas que
relacionan a dos observadores inerciales respetando la constancia de
la velocidad de la luz nos encontramos con que el tiempo deja de ser
inerte y participa de la cinem\a tica y la din\a mica.  Esta propiedad
de la Teor\ii a Especial de la Relatividad (que la relatividad general
lleva a sus \u ltimas consecuencias) obliga a abandonar no solamente
la idea de que el espacio y tiempo son estructuras independientes,
sino tambi\e n el que la geometr\ii a del continuo
espacio-temporal es euclidiana. Esto tuvo consecuencias
espectaculares sobre la estructura del universo.

Vamos a analizar con cierto detalle las consecuencias de estos dos
postulados.  Para simplificar la visualizaci\oo n geom\e trica 
consideraremos un mundo en el que nos olvidamos de una de las coordenadas
espaciales. Es decir, los puntos de nuestro espacio-tiempo en lugar de
venir etiquetados por cuatro coordenadas $(t,x,y,z)$, tres espaciales
y una temporal, vendr\a n descritos s\oo lo por tres coordenadas
$(t,x,y)$.  Esta simplificaci\oo n no presenta ning\u n problema porque
a fin de cuentas lo que queremos ilustrar es como la geometr\ii a
cambia cuando tenemos en cuenta el tiempo.  As\ii, en los diagramas
que apareceran a continuaci\oo n la coordenada vertical representar\a\
el tiempo y no la distancia espacial al plano $x$-$y$.  Asimismo, y \e
sto requiere un esfuerzo adicional, al hacer los dibujos vamos a medir
el tiempo en unidades de longitud.  Dado que la velocidad de la luz es
la misma para todos los observadores inerciales podemos utilizar como
unidad de tiempo el intervalo temporal necesario para que la luz
recorra una determinada distancia, digamos un cent\ii metro.  Esta
unidad de tiempo es igual a $(1\mbox{ cm})/(3\times 10^{10} \mbox{
cm}\mbox{ s}^{-1})
\approx 3.3 \times 10^{-11}\mbox{ s}$, 
!`esto es, unas tres cienmilmillon\e simas de segundo! 
Esto no es una complicaci\oo n gratuita. La raz\oo n para utilizar estas 
unidades de tiempo es que al hacerlo la escala en el eje $t$ est\a\
en relaci\oo n 1:1 con la escala en los ejes $x$ e $y$. Si hubieramos 
mantenido las unidades ordinarias (por ejemplo tiempo en segundos y 
distancias en cent\ii metros) los diagramas que obtendriamos ser\ii an 
bastante poco ilustrativos\footnote{Aunque medir tiempos en unidades
de longitud pueda parecer extra\nn o a primera vista, estamos sin embargo
acostrumbrados a medir distancias en unidades de tiempo. Esto es lo
que hacemos cuando expresamos las distancias entre galaxias en a\nn os-luz.}.  

Esta elecci\oo n de unidad de tiempo es equivalente a escoger un
sistema de unidades en que la velocidad de la luz toma el valor $c=1$.
A pesar de esto en las formulas seguiremos manteniendo la presencia
expl\ii cita de la velocidad de la luz como $c$.  Fijadas pu\e s las
unidades a utilizar podemos continuar con el estudio de la geometr\ii
a del espacio-tiempo. Imaginemos que nos encontramos en el centro de
nuestro sistema de coordenadas $x=0,y=0$ en el instante $t=0$ y
encendemos una bombilla. La luz formar\a\ un frente de onda circular
(esf\e rico, si estuviesemos en un espacio con tres dimensiones
espaciales) que se alejar\a\ radialmente en todas direcciones a la
velocidad de la luz. Transcurrido un tiempo $t$ desde que hemos
encendido la bombilla el radio del frente de ondas ser\a\ $ct$ y
cualquier punto sobre dicho frente con coordenadas espaciales $(x,y)$
satisfar\a\ la ecuaci\oo n
\begin{equation}\label{cono}
c^2 t^2=x^2+y^2.
\end{equation}

\begin{figure}
\centerline{\epsfbox{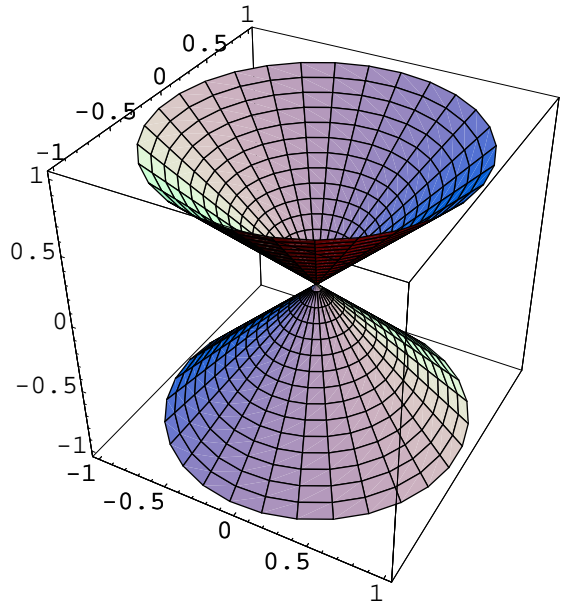}}
\caption[]{El cono de luz asociado a cualquier punto del espacio de
Minkowski utilizando nuestras unidades $c=1$.}
\label{cono-fig}
\end{figure}

Esta ecuaci\oo n la podemos visualizar dibujando la superficie que
define en el espacio-tiempo de coordenadas $(t,x,y)$, tal y como hemos
hecho en la figura \ref{cono-fig}. La figura geom\e trica obtenida se
conoce como el cono de luz. N\oo tese que \e ste tiene dos hojas; en la hoja
superior todos los puntos tienen coordenada temporal $t>0$ y por lo
tanto se encuentran en el futuro del suceso correspondiente a encender
la bombilla, que en el espacio de Minkowski corresponde al origen de
coordenadas. Por otra parte en la hoja inferior del cono $t<0$ y todos
los puntos se encuentran en el pasado del origen.

A los puntos del espacio de Minkowski $(t,x,y,z)$ se les suele
denominar ``sucesos'', ya que describen un punto de espacio $(x,y,z)$
en un determinado instante de tiempo $t$. Una de las consecuencias de
los postulados de la relatividad especial es que la velocidad de la
luz es de hecho la velocidad limite de las se\nn ales f\ii sicas. Como
consecuencia de esto el cono de luz nos permite representar adem\a s
la estructura causal del espacio-tiempo.  En la
figura \ref{cono-fig} vemos que el cono de luz divide al espacio-tiempo en dos
regiones bien definidas: por una parte los puntos dentro del cono y
sobre su superficie y por otro lado los puntos exteriores al cono.  La
clave est\a\ en darse cuenta de que los sucesos situados en el
exterior del cono de luz no pueden tener relaci\oo n causal con el
suceso situado en el origen de coordenadas. Efectivamente, si
quisieramos enviar una se\nn al entre el origen y cualquier punto
exterior al cono de luz, dicha se\nn al tendr\ii a que propagarse a
velocidad mayor que la de la luz. Como consecuencia, el suceso
correspondiente al origen no puede influir en ning\u n suceso futuro
($t>0$) localizado fuera del cono de luz, e inversamente, ning\u n
suceso con $t<0$ fuera de cono de luz puede influenciar causalmente a
lo que suceda en el origen.

Por el contrario, los puntos dentro o sobre el cono de luz est\a n en
contacto causal con el origen. Esto significa que es posible enviar
una se\nn al desde el origen a todos los puntos en el interior de la
hoja superior del cono, o que desde cualquier punto dentro de la hoja
inferior del cono es posible enviar una se\nn al que alcance el
origen. En ambos casos la se\nn al enviada se propaga a velocidad
menor o igual que la de la luz.  De esta forma podemos decir que todos
los puntos dentro o sobre la hoja inferior del cono de luz forman
nuestro pasado, en el sentido de que constituyen el conjunto de
sucesos que han podido influir el aqu\ii\ y el ahora, que corresponde
al v\e rtice del cono de luz. Por ejemplo, cuando miramos a las
estrellas lo que estamos haciendo es ver parte de la superficie pasada
(hoja inferior) de nuestro cono de luz, ya que los puntos de \e ste
corresponden a aquellos sucesos desde los cuales es posible enviar una
se\nn al luminosa al origen del sistema de coordenadas.

El cambio que esto supone con respecto a la imagen pre-einsteniana de
mundo es radical. Ahora el tiempo no s\oo lo deja de ser un par\a metro
espectador para pasar a participar de la cinem\a tica y la din\a mica,
sino que adem\a s cada suceso en el espacio-tiempo tiene asociado su
cono de luz local que determina su relaci\oo n causal con el resto de
los sucesos que forman dicho espacio-tiempo.

Al igual que el espacio eucl\ii deo, el espacio de Minkowski viene 
equipado con su geometr\ii a, esto es una forma de medir la distancia
entre dos puntos de dicho espacio. Consideremos dos sucesos $A$ y $B$
cuya diferencia de coordenadas sea $\Delta t$, $\Delta x$ y $\Delta y$. 
La ``distancia'' espacio-temporal entre ambos sucesos 
est\a\ dada entonces por
\begin{equation}\label{minkowski}
\Delta s^2=c^2 \Delta t^2-\Delta x^2-\Delta y^2\,.
\end{equation}
A diferencia del espacio eucl\ii deo, la distancia ya no est\a\ simplemente
dada por el teorema de Pit\a goras. M\a s a\u n, en el espacio de Minkowski
la cantidad $\Delta s^2$ no tiene porqu\e\ ser positiva. Su signo, como
veremos a continuaci\oo n, nos indica la posible relaci\oo n causal entre ambos
sucesos.

Consideremos de nuevo el cono de luz de la figura \ref{cono-fig} e 
identifiquemos el suceso $A$ con el origen del sistema de coordenadas 
$(0,0,0)$. En este caso tenemos tres posibilidades para el valor
del intervalo $\Delta s^2$ entre $A$ y $B$: 
\begin{itemize}
\item
$\Delta s^2>0$. En este caso el suceso $B$ se encuentra en el interior
del cono de luz. Diremos entonces que ambos sucesos est\a n separados 
por un intervalo temporal.

\item
$\Delta s^2=0$. De la propia definici\oo n del cono de luz vemos que en 
este caso el suceso $B$ yace sobre la superficie del cono. Se dice
entonces que el intervalo entre los dos sucesos es nulo. 

\item
$\Delta s^2<0$. Ahora el suceso $B$ se encuentra en el exterior del cono de
luz y el intervalo entre los sucesos $A$ y $B$ se denomina espacial.

\end{itemize}

La historia de un sistema viene representada por una curva en 
el espacio de Minkowski. Las trayectorias que describen los sistemas
f\ii sicos son curvas llamadas temporales. Esto quiere decir que
cada uno de los puntos de la curva se encuentra en el interior de todos
los conos de luz con centro en cada punto de la misma curva. \E sta es la
forma geom\e trica de decir que la velocidad de los sistemas f\ii sicos
no puede exceder la de la luz.

Para ilustrar algunos detalles interesantes de las propiedades del
espacio-tiempo, simplificaremos las gr\a ficas a\u n m\a s ignorando
tambi\e n la coordenada $y$. Por lo tanto en muchas de las figuras
nuestro espacio tiempo ser\a\ bidimensional con coordenadas $(t,x)$ y
las trayectorias de los sistemas ser\a n curvas en este plano. La
geometr\ii a local est\a\ dada por la ecuaci\oo n (\ref{minkowski})
pero suprimiendo la coordenada $y$
\be
\label{twod}
\Delta s^2=c^2 \Delta t^2-\Delta x^2\,.
\ee

\begin{figure}
\centerline{ \epsfxsize=3.0truein \epsfbox{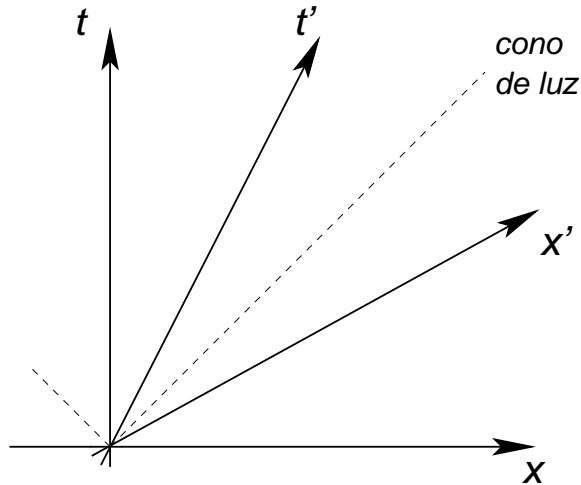}}
\caption[]{Representaci\oo n dentro de la geometr\ii a euclidiana de la
transformaci\oo n de Lorentz (\ref{lorentz})}
\label{lorentz-fig}
\end{figure}

Geom\e tricamente las transformaciones entre dos sistemas de referencia
inerciales corresponden a cambios de coordenadas $(t,x)\rightarrow
(t',x')$ que mantienen invariante el intervalo espacio-temporal
(\ref{twod}) entre dos sucesos arbitrarios\footnote{El an\a logo en el
espacio eucl\ii deo ser\ii an las transformaciones ortogonales que
dejan invariante la distancia entre dos puntos cualequiera de dicho
espacio.}. Estas
transformaciones se conocen como transformaciones de
Lorentz. Imaginemos que $(t',x')$ corresponde a las coordenadas de un
suceso medidas en un sistema de referencia inercial que se mueve con
respecto al sistema $(t,x)$ con velocidad $v$ a lo largo del eje
$x$. No es dificil verificar que las transformaciones
\begin{equation}\label{lorentz}
x'\, = \, {{x - v t} \over \sqrt{1-{v^2/ c^2}}}\,, \qquad t'\, = \, 
{{t - {v x/ c^2}} \over \sqrt{1-{v^2/ c^2}}},
\end{equation}
dejan el intervalo (\ref{twod}) invariante y se reducen a las
transformaciones de Galileo cuando la velocidad $v$ es mucho menor que
la de la luz. Si representamos su efecto en un plano eucl\ii deo el
resultado es bastante chocante (v\e ase figura
\ref{lorentz-fig}). N\oo tese sin embargo que, al dejar invariante
el intervalo (\ref{twod}) las transformaciones de Lorentz dejan 
invariante el cono de luz y la estructura causal del espacio-tiempo.

\section{La paradoja de los gemelos}

Una de las paradojas cl\a sicas que se presentan casi siempre en
exposiciones elementales de la teor\ii a de la relatividad es la
llamada la paradoja de los gemelos. Supongamos dos gemelos id\e
nticos, uno de los cuales inicia un viaje de ida y vuelta con velociad
$v$ mientras que el otro se queda tranquilamente esperando el regreso
de su hermano. En el sistema de referencia del gemelo que espera el
reloj de su hermano viajero ir\a\ m\a s despacio que el suyo y por lo
tanto en el reencuentro ser\a\ m\a s joven que \e l. Pero si uno toma el
sistema de referencia que viaja con el segundo gemelo, es el primer
gemelo el que se mueve con velocidad $-v$ y por lo tanto cuando
vuelvan a verse el hermano que se qued\oo\ esper\a ndo habr\a\ de ser
el m\a s joven. He aqu\ii\ la paradoja, ya que cuando los dos hermanos
vuelvan a verse las dos alternativas son mutuamente excluyentes.

Si pensamos de forma espacio-temporal, veremos que esta paradoja no es
tal.  Pero antes necesitamos entender un poco mejor el significado de
$\Delta s^2$.  Si en la ecuaci\oo n (\ref{twod}) en lugar del signo
negativo los dos t\e rminos en el miembro de la derecha tuviesen un
signo positivo podr\ii amos interpretar $\Delta s$ como una
distancia. Pero si $\Delta s^2$ no tiene un signo definido ?`c\u al es
entonces la interpretaci\oo n del intervalo? En el caso en que $\Delta
s^2>0$ el intervalo admite una interpretaci\oo n f\ii sica muy
interesante. En este caso podemos considerar un observador inercial
que se mueve con respecto a nuestro sistema de referencia y cuya
trayectoria en el espacio de Minkowski pasa por los sucesos $A$ y
$B$. Para este observador ambos sucesos ocurren en el origen de su
sistema de referencia ($\Delta x=0$) pero separados por un intervalo
de tiempo $\Delta \tau$. Dado que el intervalo $\Delta s^2$ entre los
dos sucesos es el mismo para todos los observadores inerciales tenemos
que
\be\label{tiempop}
\Delta\tau^2 = {\Delta s^2\over c^2}.
\ee
En este caso el intervalo entre los dos sucesos $A$ y $B$ est\a\ directamente
relacionado con el tiempo propio de un observador para el que ambos sucesos
ocurren en el mismo punto del espacio (su origen de coordenadas, por ejemplo).

\begin{figure}
\centerline{ \epsfxsize=3.0truein \epsfbox{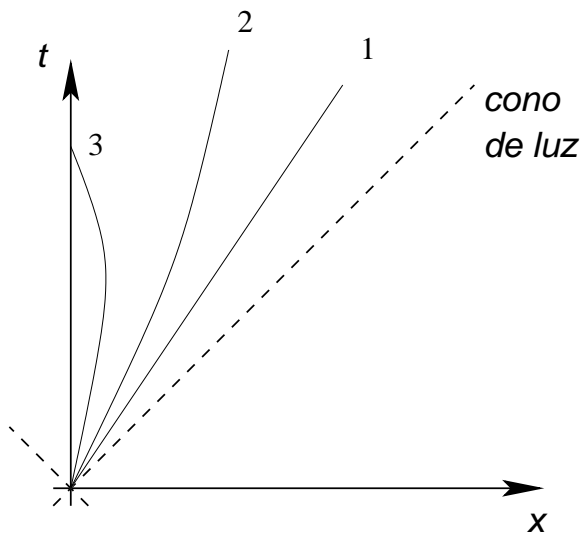}}
\caption[]{Varias trayectorias en el espacio de Minkowski.}
\label{4}
\end{figure}

De hecho el concepto de tiempo propio de un observador en movimiento
con respecto a un sistema de referencia puede generalizarse al caso
en que dicho observador se mueve con velocidad variable. Como hemos
comentado m\a s arriba, los sistemas f\ii sicos se representan en el
espacio de Minkowski como curvas temporales. En la figura \ref{4} se
muestran algunos ejemplos de dichas curvas en el plano $x$-$t$. 
Por ejemplo, la curva correspondiente a un observador en reposo 
situado en $x=0$ simplemente coincide con el eje temporal. Por otra parte
si el observador se mueve con velocidad uniforme con respecto al
sistema de referencia su trayectoria en el espacio de Minkowski ser\a\ 
un recta como la l\ii nea 1. La curva 2 corresponde
a un observador que se mueve con velocidad variable, 
mientras que en el caso de la curva 3 tendremos un observador cuyo
viaje empieza y termina en el origen, donde se encontrar\a\ de nuevo
con el observador en reposo. 

Para los observadores inerciales sabemos que el tiempo propio
trascurrido entre dos sucesos se relaciona con el intervalo
espacio-temporal entre ambos por la f\oo rmula (\ref{tiempop}). Esta
relaci\oo n entre tiempo propio y ``distancia'' espacio-temporal puede
extenderse f\a cilmente al caso de observadores no
inerciales. Consideremos de nuevo la trayectoria 2 en la figura
\ref{4}. Siempre podemos ``rectificar'' dicha curva con precisi\oo n
arbitraria por una l\ii nea poligonal de forma que en cada segmento de
dicha curva la relaci\oo n (\ref{tiempop}) entre el tiempo propio y el
intervalo entre los extremos del segmento es aplicable.  Sumando los
distintos valores de $\Delta \tau$ en cada segmento (esto es, haciendo
la integral sobre $\tau$ a lo largo de la trayectoria) obtenemos el
tiempo que marcar\a\ el reloj que acompa\nn a al observador en su
recorrido.  Por lo tanto para un observador arbitrario su tiempo
propio estar\a\ dado por (\ref{tiempop}), d\oo nde ahora $\Delta s$ es
la ``longitud'' de la trayectoria de dicho observador en el espacio
de Minkowski. Es importante notar que en este an\a lisis es crucial el
que la curva sea temporal, ya que esto garantiza que $\Delta s$ en
cada segmento es una cantidad real.

Sabiendo esto podemos intentar clarificar la paradoja de los gemelos.
Imaginemos un grupo de gemelos que comienzan su viaje de 
ida y vuelta en $x=0$ con diferentes velocidades (v\e ase la figura 5),
mientras que uno de ellos (el observador 1) se queda en $x=0$ viendo
pasar el tiempo.
\begin{figure}
\centerline{ \epsfxsize=3.0truein \epsfbox{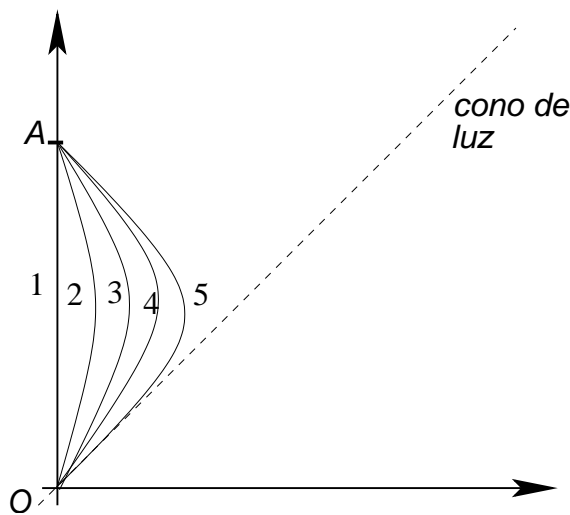}}
\caption[]{Familia de observadores considerados en el texto y donde
se compara la juventud relativa de ellos en el suceso A.}
\label{5}
\end{figure}
Dado que las trayectorias espacio-temporales de los diferentes
observadores tendr\a n diferentes longitudes nos podemos preguntar
cu\a l de los gemelos ser\a\ el m\a s joven cuando vuelvan a encontrarse
en $A$. Si nos dej\a semos llevar por nuestra intuici\oo n euclidiana
y concluy\e semos que la trayectoria 5 es la de mayor longitud,
llegar\ii amos a la conclusi\oo n err\oo nea de que el observador 5
ser\ii a el m\a s viejo al llegar a $A$.  Sin embargo, precisamente
debido al signo menos en la f\oo rmula (\ref{twod}), la trayectoria 5 es
la de menor longitud espacio-temporal y por lo tanto ser\a\ para este
observador para el que el viaje haya durado menos. Una manera
intuitiva de comprobar este hecho sin hacer ning\u n c\a lculo es
darse cuenta de que esta trayectoria es la m\a s cercana al cono de
luz y que los puntos sobre este satisfacen $\Delta s^2=0$. Por lo
tanto en la fiesta de reencuentro que los gemelos organizar\a n en $A$
el gemelo 1 ser\a\ m\a s viejo que el gemelo 2, \e ste a su vez ser\a\
m\a s viejo que el gemelo 3 y as\ii\ sucesivamente, siendo el gemelo 5
el m\a s joven de todos.

De nuestro an\a lisis anterior se sigue que el gemelo que viaja ser\a\
siempre m\a s joven que el que se qued\oo\ en reposo en un sistema de
referencia inercial. Lo que elimina la paradoja es que el sistema
de referencia del gemelo que viaja no es un sistema inercial y por lo
tanto no podremos aplicar las f\oo rmulas de la relatividad especial para
concluir que respecto de dicho sistema de referencia el reloj 
del gemelo ``inercial'' va m\a s despacio.

Finalmente, como corolario a la figura \ref{5}, nos gustar\ii a se\nn
alar que seg\u n la Teor\ii a Especial de la Relatividad cuando
buscamos la trayectoria de una part\ii cula libre entre un suceso
inicial y otro final, la din\a mica selecciona aquella que, pasando
por los dos sucesos, corresponde a un tiempo propio mayor.

\section{La relatividad general}

En su art\ii culo de 1905 \cite{1905} Einstein hizo compatibles la
mec\a nica y el electromagnetismo en beneficio de \e ste \u ltimo: las
ecuaciones de Maxwell toman la misma forma en todos los sistemas de
referencia inerciales y la velocidad de la luz es la misma para todos
los observadores, proporcionando adem\a s la velocidad m\a xima para
la transmisi\oo n de informaci\oo n o interacciones. Cada suceso est\a\
equipado con su cono de luz y (\ref{twod}) define la nueva geometr\ii
a local.  

Haciendo un inciso, es francamente curioso que la geometrizaci\oo n de
la relatividad especial llevada a cabo por Minkowski no fue del agrado
de Einstein que la tild\oo{} de ``erudici\oo n innecesaria''
\cite{Pais}. Sin embargo, como \e l mismo tuvo que admitir m\a s adelante, 
la formulaci\oo n minkowskiana parec\ii a la m\a s adecuada para asaltar el 
problema que el propio Einstein se hab\ii a propuesto poco tiempo despu\e s de
formular la Teor\ii a Especial de la Relatividad: la formulaci\oo n de una
teor\ii a relativista de la gravitaci\oo n. Este era el paso l\oo gico 
despu\e s de que la relatividad especial hubiese establecido la armon\ii a
entre la mec\a nica y la electrodin\a mica.

Al igual que con la Teor\ii a Especial de la Relatividad, el camino de
Einstein hacia la formulaci\oo n de la relatividad general se inicia,
en 1907, con un experimento imaginario: para el infortunado individuo
que cae desde lo alto del tejado de una casa el campo gravitatorio
deja de existir mientras dura su caida. Durante ese tiempo el
observador experimentar\a{} la misma sensaci\oo n que tendr\ii a si se
encontrara en ingravidez. Esta ``simple'' observaci\oo n, que Einstein
calific\oo{} como ``la ocurrencia m\a s afortunada de mi vida'',
constituye el llamado principio de equivalencia y result\oo{} ser la
piedra angular sobre la que se construy\oo{} todo el edificio de la
Teor\ii a General de la Relatividad.

Para entender el principio de equivalencia un poco mejor vamos a
considerar un ejemplo familiar. Todos estamos ya acostumbrados a ver
las im\a genes de astronatuas en estaciones espaciales que orbitan
alrededor de la Tierra y en las que estos flotan como si se
encontrasen en ausencia de gravedad.  Sin embargo un sencillo calculo
muestra que la aceleraci\oo n de la gravedad terrestre en una estaci\oo
n espacial que se encuentra a una altura de, por ejemplo, $h=400$
kil\oo metros sobre la superficie de la Tierra no es ni mucho menos
despreciable: $$ g=9.8\left(1+{h\over R_{\oplus}}\right)^{-2}\mbox{ m
}\mbox{s}^{-2} \simeq 8.7 \mbox{ m }\mbox{s}^{-2}\,.  $$ Por lo tanto
la aceleraci\oo n de la gravedad all\a{} arriba es s\oo lo ligeramente
m\a s peque\nn a que la que experimentamos sobre la superficie de la
Tierra.  ?`C\oo mo podemos explicar entonces que los astronautas
parezcan no sentir ning\u n tipo de atracci\oo n gravitacional?  Lo que
realmente ocurre es que tanto los astronautas como el equipo y la
propia nave al orbitar est\a n en caida libre hacia la Tierra y por lo
tanto de acuerdo con el principio de equivalencia es como si no
sintiesen el campo gravitatorio.

Igualmente un astronauta en una nave espacial localizada muy lejos de 
cualquier cuerpo celeste y que por lo tanto se encuentra en estado de 
ingravidez puede simular el campo gravitatorio terrestre usando 
el principio de equivalencia. Si este astronauta enciende los cohetes de
su nave de forma que \e sta adquiera una aceleraci\oo n igual a $9.8 \mbox{ m }
\mbox{s}^{-2}$ los tripulantes de dicha nave experimentar\a n una fuerza
en direcci\oo n contraria que dar\a\ lugar a una aceleraci\oo n en todos 
los cuerpos igual a $9.8 \mbox{ m }\mbox{s}^{-2}$ independientemente de su
masa.

El principio de equivalencia supuso para Einstein el inicio del largo
y tortuoso camino que tras ocho a\nn os y m\a s de un fracaso
le condujo a la descripci\oo n del campo gravitatorio como un efecto de
la geometr\ii a del espacio-tiempo. Como hemos indicado m\a s arriba,
todos los observadores inerciales observan la misma m\e trica del
espacio-tiempo. Esto no es cierto sin embargo para observadores no
inerciales (acelerados) para los cuales la m\e trica vendr\a{} dada
por una forma cuadr\a tica general. El principio de equivalencia nos
dice que, localmente, podemos eliminar un campo gravitatorio
simplemente dejandonos caer con \e l o crearlo acelerando nuestro
sistema de referencia inercial.

Ya hemos mencionado que en la geometr\ii a riemanniana el teorema de 
Pit\a goras gobierna la geometr\ii a local.  En un entorno suficientemente 
peque\nn o de un punto dado la geometr\ii a euclidiana puede aplicarse 
con un alto grado de precisi\oo n. Este hecho nos da la pista para 
concretar el principio de equivalencia en t\e rminos geom\e tricos.
Lo que la observaci\oo n del individuo que cae del tejado nos est\a\
diciendo es que en presencia de la gravedad existen sistemas de
referencia en los que la geometr\ii a del espacio-tiempo es localmente la 
de Minkowski. As\ii\ elementos como el cono de luz y la estructura
causal que hemos discutido m\a s arriba siguen formando parte fundamental
de la descripci\oo n de la naturaleza cuando tenemos en cuenta el efecto
de los campos gravitatorios.

La imagen del mundo que surge de la Teor\ii a General de la
Relatividad es la de un espacio-tiempo riemanniano d\oo nde localmente la
geometr\ii a es Minkowskiana. Junto a esto, los sistemas de referencia
inerciales dejan de tener el papel fundamental que jugaban en la
relatividad especial; como nos ense\nn a el principio
de equivalencia el efecto de la gravedad es localmente equivalente al
de las fuerzas de inercia que aparecen en sistemas de referencia no
inerciales y por lo tanto estos han de considerarse al mismo nivel que
los sistemas de referencia inerciales. Como consecuencia,
la m\e trica del espacio-tiempo no tendr\a\ en general la forma tan simple que
habiamos usado en la ecuaci\oo n (\ref{minkowski}). Ahora el intervalo entre
dos sucesos pr\oo ximos vendr\a\ determinado por diez coeficientes que 
constituyen el llamado tensor m\e trico
\begin{eqnarray}\label{metrica}
\Delta s^2 = \sum_{i,j=0}^3 g_{ij}\, \Delta x^i \, \Delta x^j \,.
\end{eqnarray}
Aqu\ii{} hemos etiquetado las coordenadas espacio-temporales como
$(x^0,x^1,x^2,x^3)$ y los diez coeficientes independientes $g_{ij}$
($g_{ij}= g_{ji}$) son en general funcion de las coordenadas. Lo
importante es que localmente siempre existe un sistema de referencia
con coordenadas $(x'^0,x'^1,x'^2,x'^3)$ tales que en dicho sistema
$\Delta s^2$ toma la forma
\begin{eqnarray}
\Delta s^2=(\Delta x'^0)^2-(\Delta x'^1)^2-(\Delta x'^2)^2-(\Delta x'^3)^2\,.
\end{eqnarray}
\E sta es la expresi\oo n matem\a tica del principio de 
equivalencia.

A pesar de la mayor complejidad matem\a tica del intervalo
(\ref{metrica}) su interpretaci\oo n f\ii sica sin embargo es
completamenta an\a loga al caso minkowskiano.
Al igual que en la ecuaci\oo n (\ref{tiempop}) $\Delta\tau = \Delta s/c$ 
define localmente el tiempo propio de un observador en movimiento arbitrario.
Asimismo cada suceso tiene su cono de luz asociado, que una vez m\a s se define
como el lugar geom\e trico de los sucesos para los que $\Delta s^2=0$. 
Es decir, el cono de luz est\a\ definido otra vez por las trayectorias de 
propagaci\oo n libre de los rayos de luz. Las trayectorias de las 
part\ii culas se obtienen maximizando el tiempo propio, al igual que en el
espacio de Minkowski. Las curvas que unen dos sucesos y a lo largo de las
cuales el tiempo propio es m\a ximo se conocen como geod\e sicas. Por lo 
tanto, en presencia de un campo gravitacional, las part\ii culas describir\a n
geod\e sicas en un cierto espacio-tiempo curvado cuya geometr\ii a est\a\
descrita por (\ref{metrica}).

Este argumento heur\ii stico nos da la pista sobre c\oo mo describir el
campo gravitacional en t\e rminos geom\e tricos. Sin embargo no nos
dice nada sobre la din\a mica del campo gravitacional, esto es, sobre
c\oo mo la distribuci\oo n de materia determina la
gravedad. \E ste fue de hecho el gran problema al que tuvo que
enfrentarse Einstein durante los a\nn os que dur\oo{} la elaboraci\oo n
de la Teor\ii a General de la Relatividad y que absorbi\oo\ la mayor
parte de su actividad llev\a ndole casi hasta la extenuaci\oo n f\ii
sica.

No vamos a entrar aqu\ii\ a explicar en detalle las ecuaciones de la
relatividad general. Pero dada su belleza no podemos resistirnos a
escribir las ecuaciones de Einstein del campo gravitatorio
\cite{rmunu}
\begin{eqnarray}\label{einstein}
R_{ij}-{1\over 2}g_{ij}R={8\pi G\over c^3}\,T_{ij} \,.
\end{eqnarray}
El miembro de la izquierda de esta ecuaci\oo n contiene
\u nicamente cantidades asociadas con la geometr\ii a del espacio 
tiempo: $R_{ij}$ es el llamado tensor de Ricci que se expresa en 
t\e rminos de la m\e trica y sus derivadas segundas y $R$ es el escalar 
de curvatura, definido como la traza del tensor de Ricci 
\begin{eqnarray}
R=\sum_{i,j=0}^{3}g^{ij}R_{ij}\,.
\end{eqnarray}
Por otra parte el miembro de la derecha de (\ref{einstein}) 
contiene la informaci\oo n sobre la distribuci\oo n de materia-energ\ii a
del espacio tiempo, codificada en el tensor de energ\ii a-momento $T_{ij}$.
Vemos por lo tanto que, moralmente, las ecuaciones de Einstein las
podemos escribir como
$$
\mbox{Geometr\ii a = Materia.}
$$ 
Es decir, la presencia de materia tiene como resultado la curvatura de
espacio-tiempo, algo que nosotros observamos en forma de campos
gravitatorios. Adem\a s, la teor\ii a de Newton de la gravitaci\oo n se obtiene
entonces de las ecuaciones de Einstein en el l\ii mite de bajas
velocidades y peque\nn as concentraciones de masa.

\section{Agujeros negros}

Desgraciadamente no nos es posible analizar en detalle todas y cada
una de las fascinantes predicciones de la relatividad general, muchas
de las cuales ha sido comprobadas observacionalmente con precisiones
asombrosas. Aqu\ii\ mencionaremos las que son quiz\a s las m\a s conocidas:
la existencia de agujeros negros y la expansi\oo n del Universo. 

\begin{figure}
\centerline{ \epsfxsize=6.0truein \epsfbox{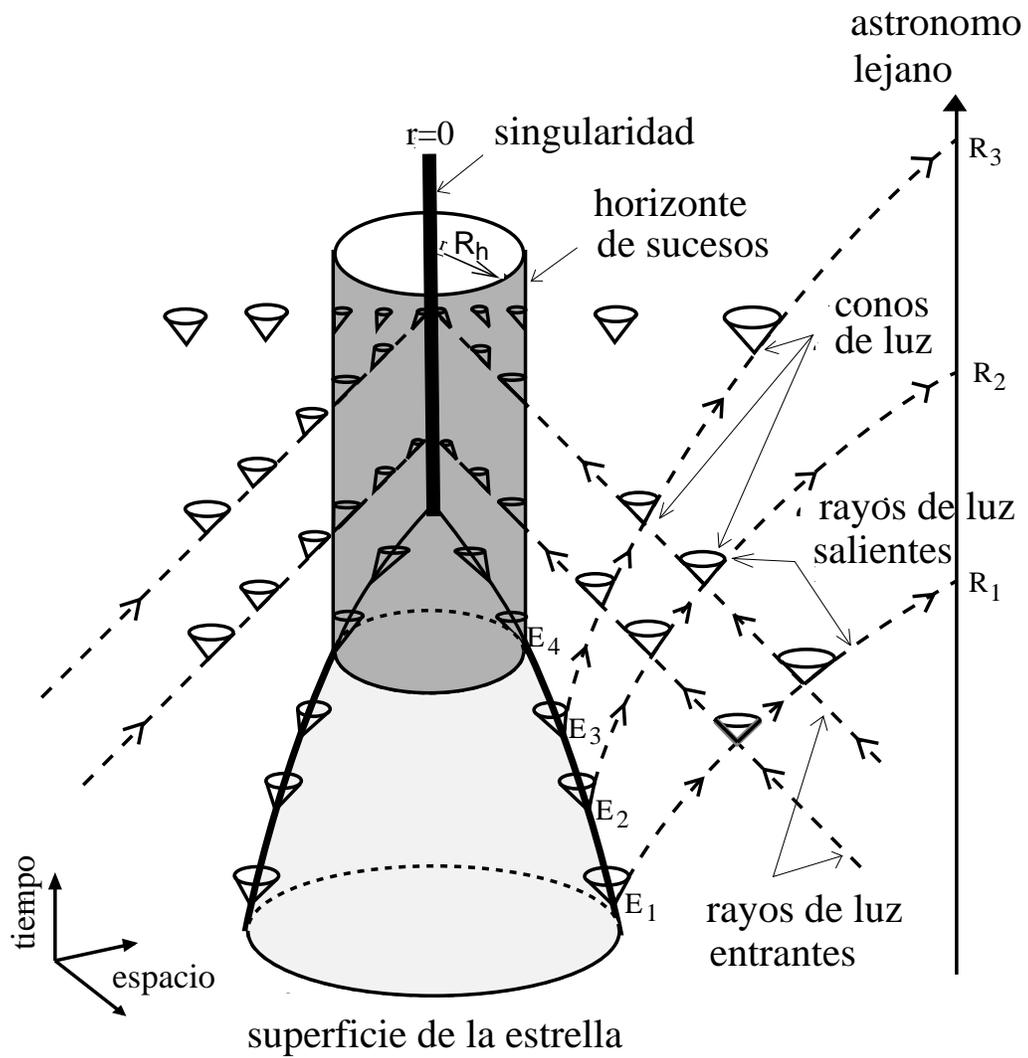}}
\caption[]{Estrucutra causal en el espacio-tiempo alrededor de
una estrella que colapsa \cite{luminet}.}
\label{schw}
\end{figure}

Uno de los resultados m\a s profundos de la relatividad general es que
en cierto sentido la teor\ii a conoce sus l\ii mites.  Bajo ciertas
condiciones muy generales la materia colapsa sin l\ii mite hasta
formar una singularidad en el espacio-tiempo.  El ejemplo m\a s t\ii pico 
de este fen\oo menos es de una estrella muy masiva (cuya masa,
digamos, es muchas veces la del Sol) que colapsa bajo la influencia de
su propia gravedad\footnote{Un problema f\ii sico muy interesante es
el estudio de los estadios finales de la vida de una estrella y sus
diferentes destinos dependiendo de su masa: enanas blancas, 
estrellas de neutrones o agujeros negros.}. En lugar de presentar
un an\a lisis detallado de este proceso en t\e rminos matem\a ticos
vamos a intentar entender los aspectos esenciales de la formaci\oo n de
un agujero negro con la ayuda de la figura
\ref{schw}.
En este dibujo tambi\e n puede verse la orientaci\oo n de los conos de
luz en cada punto. Recordemos que como las propiedades geom\e tricas
del espacio-tiempo en torno a la estrella varia de un punto a otro lo
mismo ocurre con la estructura de conos de luz.  En particular estos
apuntar\a n en diferentes direcciones en cada punto, a diferencia de lo
que ocurre en el espacio de Minkowski d\oo nde todos los conos de luz
son paralelos.

Imaginemos que inicialmente nos encontramos en el extremo inferior de
la figura sobre la superficie de la estrella.  A medida que \e sta
colapsa los ejes de los conos de luz sobre los puntos en la superficie
de la estrella se van inclinando m\a s y m\a s hacia el interior de la
estrella. De hecho llega un momento en que los rayos de luz emitidos
desde la superficie de la estrella no pueden escapar a regiones
lejanas. Este fen\oo meno corresponde a la formaci\oo n del llamado
horizonte de sucesos, y tiene lugar cuando el radio de la estrella
colapsante decrece por debajo del radio de Schwarzschild\footnote{
Para hacernos una idea del tama\nn o de este radio baste decir que
para un objeto con la masa de la Tierra, $R_{\rm h}$ ser\ii a
approximadamente igual a 9 mil\ii metros.}
\begin{eqnarray}
R_{\rm h}={2 G M \over c^2}\,,
\end{eqnarray}
d\oo nde $M$ es la masa de la estrella.  En la figura \ref{schw} el horizonte
de sucesos corresponde a la superficie cil\ii ndrica de radio $R_{\rm
h}$. Como vemos, los conos de luz sobre el horizonte son tangentes
esta superficie y apuntan hacia el interior de ella. Como consecuencia
ninguna se\nn al emitida sobre o dentro del horizonte podr\a\ alcanzar
al astronauta que observa el colapso desde la lejan\ii a.

Una vez formado el horizonte la estrella seguir\a\ su proceso de
colapso hasta que en un tiempo finito, medido por el observador sobre
su superficie, toda la masa de \e sta se concentre en el punto $r=0$
formando una singularidad en el espacio-tiempo d\oo nde la curvatura
tiende a infinito. Dado que ninguna se\nn al puede abandonar el
horizonte, esta singularidad no ser\a\ observable desde el exterior
del agujero negro y la f\ii sica fuera del horizonte est\a\
completamente desacoplada de lo que ocurre en el interior. Para un
observador externo el agujero negro es, cl\a sicamente, una superficie
esf\e rica completamente negra de radio $R_{\rm h}$ que se traga todo
aquello que la atraviesa\footnote{De hecho, como demostr\oo\ Stephen
Hawking en 1975, debido a los efectos cu\a nticos la superficie del
agujero negro no es completamente negra, sino que emite radiaci\oo n
t\e rmica con temperatura $T_{\rm H}=\hbar c/(4\pi k_{\rm B} R_{\rm h})$,
d\oo nde $k_{\rm B}$ es la constante de Boltzmann.}. Es
interesante que en el caso que hemos discutido la presencia del
horizonte nos protege de la p\e rdida de predictibilidad que
supondr\ii a la existencia de una singularidad ``desnuda''.  En los
a\nn os setenta Roger Penrose formul\oo\ la conjetura conocida como
``censura c\oo smica". Segun ella, todo colapso gravitatorio realista
deber\ii a siempre producir un horizonte que separe a los observadores
asint\oo ticos de la singularidad. Hasta ahora, sin embargo, no se ha
dado ninguna demostraci\oo n de esta conjetura e incluso hay dudas
sobre su validez general.

\begin{figure}
\epsfxsize=11cm
\centerline{ \epsfxsize=6.0truein \epsfbox{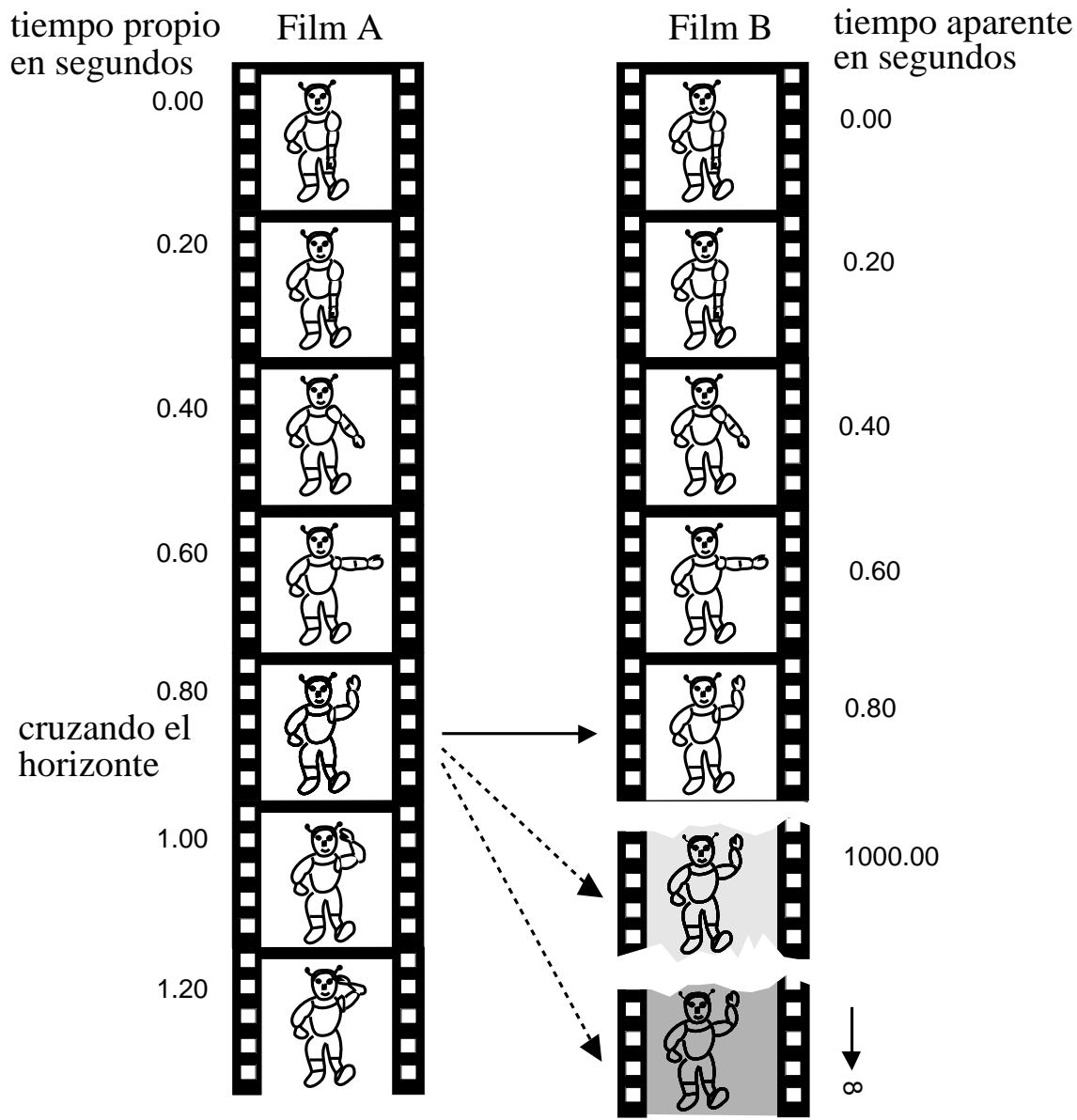}}
\caption[]{Comparaci\oo n entre el tiempo propio de un astronauta
que cae en un agujero negro (film A) y el del astr\oo nomo que observa
preocupado (film B) \cite{luminet}.}
\label{flatt}
\end{figure}

Para mejor comprender la deformaci\oo n que sobre el espacio-tiempo
produce una estrella que colapsa vamos a comparar el tiempo propio
de un astronauta incauto que se encuentra sobre la superficie de la
estrella con el tiempo que marca el reloj de otro astronauta, m\a s
precavido, que observa el colapso desde lejos. Supongamos que el astronauta
desde la superficie de la estrella nos env\ii a un saludo. La figura
\ref{flatt} nos muestra la sucesi\oo n de acontecimientos tanto para 
el astronauta que cae como para su compa\nn ero que le observa desde
lejos. 

Al principio del colapso los tiempos propios de ambos astronautas son
casi iguales, como puede verse del hecho de que los primeros
fotogramas en las dos pel\ii culas de la figura \ref{flatt} son
practicamente id\e nticos. Esto cambia, sin embargo, cuando el radio
de la estrella en colapso se acerca a $R_{\rm h}$. Como podemos ver en
los sucesos $E_{1}$, $E_{2}$ y $E_{3}$ de la figura \ref{schw}, al
acercarnos al horizonte de sucesos los conos de luz se inclinan hacia
el interior de \e ste, lo que hace que los rayos de luz emitidos por el
infortunado astronauta tarden cada vez m\a s en llegar a su compa\nn
ero. Adem\a s de tardar m\a s en llegar a su destino, las se\nn ales
luminosas enviadas desde la superficie de la estrella pierde gran
parte de su energ\ii a para poder escapar del pozo de potencial
gravitacional en el que cae el astronauta, es decir, presentar\a n un
desplazamiento hacia el rojo al ser recibidas. Esto significa que la
longitud de onda de la se\nn al que recibe el astronauta lejano ser\a\
mucho mayor que la que ten\ii a dicha se\nn al al ser emitida sobre la
superficie de la estrella.

Finalmente, cuando el astronauta atraviesa el horizonte (el suceso
$E_{4}$ en la figura \ref{schw}) su cono de luz se encontrar\a\ dentro
del horizonte por lo que le ser\a\ imposible cualquier comunicaci\oo n
con el exterior del agujero negro. Sin embargo el astronauta que
observa todo desde lejos no dejar\a\ nunca de ver a su compa\nn ero. A
medida que \e ste se acerca al horizonte la imagen del intr\e pido
astronauta se ir\a\ ralentizando hasta que finalmente quedar\a\ s\oo lo
una ``imagen fija'' que se ira haciendo m\a s y m\a s roja y d\e bil
hasta acabar por desvanecerse en un tiempo infinito. Para el
astronauta que cae, sin embargo, nada especial ocurre cuando atraviesa
el horizonte, tal y como podemos ver en la figura \ref{schw}.  Ignorante
de su destino, \e l seguir\a\ saludando a su colega hasta que al cabo
de un tiempo finito ser\a\ destruido en la singularidad.

\section{Cosmolog\ii a}

Poco despu\e s de formular la Teor\ii a General de la Relatividad en
1915 Einstein intent\oo\ su aplicaci\oo n al estudio del Universo en su
conjunto. Para ello introdujo lo que hoy conocemos como el Principio
Cosmol\oo gico, seg\u n el cual ning\u n observador en el Universo
ocupa una posici\oo n privilegiada de forma que las propiedades geom\e
tricas del Universo deben de ser independientes del punto. En general
la Teor\ii a General de la Relatividad junto con el Principio Cosmol\oo
gico conduce a un universo din\a mico. Sin embargo, cuando
Einstein public\oo\ en 1917 el que puede considerarse como el art\ii
culo fundacional de la cosmolog\ii a relativista \cite{lambda} no
hab\ii a evidencia alguna de que el Universo se estuviese expandiendo.
Por ello Einstein modific\oo\ las ecuaciones de la relatividad general
a\nn adiendo un t\e rmino adicional $\Lambda g_{ij}$ en el miembro de
la izquierda de (\ref{einstein}). Asumiendo que las secciones
espaciales del Universo eran esferas de tres dimensiones, exist\ii a
un valor de $\Lambda$ para la cual el Universo era est\a tico.

En 1929 Edwin Hubble obtuvo la primera evidencia observacional de la
expansi\oo n del Universo mediante la observaci\oo n del deplazamiento
hacia el rojo de las galaxias lejanas, que mostraba que \e stas se
alejaban de nosotros en todas las direcciones con velocidades de
recesi\oo n proporcionales a su distancia (ley de Hubble). Esto supuso
el abandono del modelo est\a tico de Einstein, as\ii\ como de la
constante cosmol\oo gica, cuya introducci\oo n el propio Einstein
calific\oo\ como ``el mayor error de mi vida''\footnote{Recientes
observaciones cosmol\oo gicas muestran, no obstante, una aceleraci\oo n
en la expasi\oo n del Universo que puede atribuirse a un valor no nulo
de la constante cosmol\oo gica.}.

Soluciones no est\a ticas compatibles con el Principio Cosmol\oo gico
fueron obtenidas por Willem de Sitter, Alexander Friedmann, Georges
Lema\^{\i}tre, Howard Robertson y Arthur Walker en los a\nn os
siguientes a la formulaci\oo n de la relatividad general. En particular
el llamado modelo de Friedmann-Lema\^{\i}tre-Robertson-Walker describe
un universo en expansi\oo n en el que las galaxias lejanas satisfacen
la ley de Hubble. Una de las predicciones m\a s importante de este
modelo es que en el pasado el Universo fue mucho m\a s denso y m\a s
caliente de lo que es en la actualidad, hasta llegar a un instante en
el que la densidad de energ\ii a y la curvatura del espacio-tiempo
divergen. Esta singularidad inicial, en la cual la teor\ii a pierde
toda predictibilidad, es el llamado {\it Big-Bang}.  Vemos que, al
igual que ocurre en el proceso de colapso gravitatorio de una estrella
supermasiva, la relatividad general predice la existencia de
singularidades en las que la propia teor\ii a deja de ser v\a lida. 
\E ste es un nuevo ejemplo de algo que ya hemos mencionado m\a s arriba:
la Teor\ii a General de la Relatividad nos proporciona sus propios
l\ii mites. En los a\nn os sesenta Roger Penrose y Stephen Hawking
demostraron rigurosamente que, bajo condiciones muy generales, la
singularidad inicial es inescapable en la cosmolog\ii a relativista.

La cosmolog\ii a de Big-Bang ha recibido muchas e importantes
comprobaciones experimentales por lo que hoy se habla de ella como del
modelo cosmol\oo gico est\a ndar. Quiz\a s la m\a s espectacular ha
sido el descubrimiento en 1964 de la radiaci\oo n de fondo de
microondas por Arno Penzias and Robert Wilson, una predicci\oo n
inequivoca del modelo de Big-Bang. Esta radiaci\oo n es una reliquia de
los tiempos inmediatamente posteriores al Big-Bang en los que el
Universo era m\a s denso y m\a s caliente.

En cosmolog\ii a, al igual que vimos al estudiar la formaci\oo n de un
agujero negro, la geometr\ii a del espacio-tiempo condiciona la
estructura causal del mismo. En t\e rminos generales sabemos que al
observar el Universo estamos de hecho explorando la superficie pasada
de nuestro cono de luz. En un espacio-tiempo eterno, como
es el espacio de Minkowski, este cono de luz se extiende
indefinidamente en el pasado por lo que, en principio, tenemos la
posibilidad de recibir se\nn ales luminosas emitidas desde regiones
arbitrariamente alejadas de nuestra posici\oo n.  Esto sin embargo no
es posible en un universo como el nuestro que se origin\oo\ en alg\u n
instante de tiempo en el pasado. La existencia del Big-Bang implica
que el Universo s\oo lo ha existido durante un tiempo finito en el pasado
y como consecuencia la luz procedente de regiones suficientemente
lejanas de nosotros no habr\a\ tenido tiempo de alcanzarnos
desde que el Universo comenz\oo\ a existir. 

\begin{figure}
\epsfxsize=11cm
\centerline{ \epsfxsize=5.0truein \epsfbox{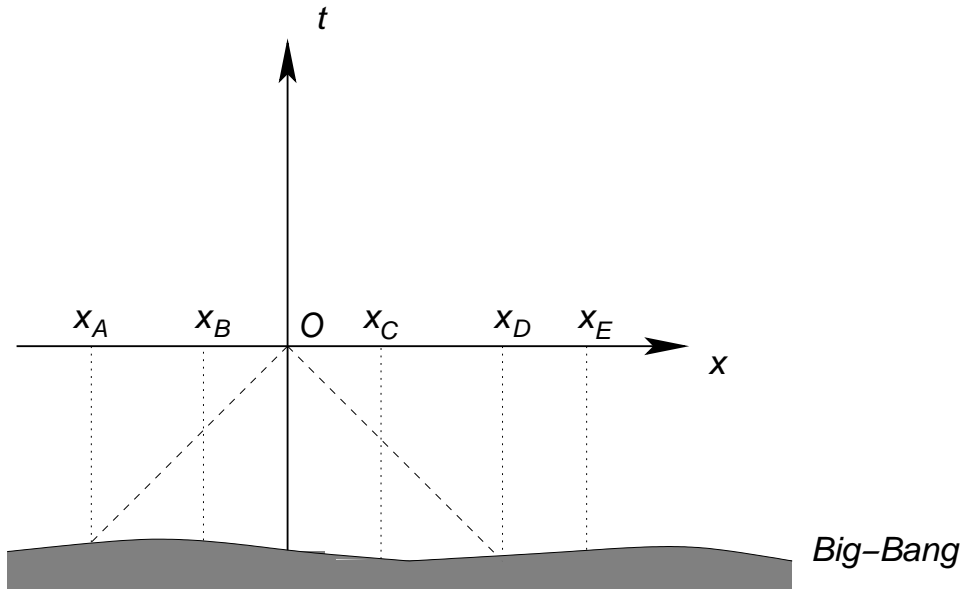}}
\caption[]{Ilustraci\oo n de la existencia de horizontes de part\ii culas.
Ning\u n suceso con $x<x_A$ o $x>x_E$ ser\a\ observables desde $O$.}
\label{horizonte}
\end{figure}

Este fen\oo meno aparece ilustrado en la figura \ref{horizonte} usando 
el espacio-tiempo bidimensional que introdujimos m\a s arriba.
Vemos que debido a la extensi\oo n finita en el pasado del cono de luz,
en $O$ s\oo lo podremos recibir se\nn ales procedentes de puntos con
coordenadas $x_{A}<x<x_{D}$. Por lo tanto cualquier suceso pasado que
haya tenido lugar m\a s all\a\ de esta regi\oo n (digamos, por ejemplo,
en un punto con coordenada espacial $x_E$) no ser\a\ accesible desde
$O$. Esto define lo que se conoce como un horizonte de part\ii culas.
Su existencia implica que aun cuando pudieramos construir instrumentos
de observaci\oo n infintamente precisos nuestras posibilidades de
observar el Universo est\a n limitadas a una regi\oo n finita alrededor
de nuestra posici\oo n.  La distancia al horizonte es aproximadamente
igual a la distancia que la luz ha tenido tiempo de recorrer desde el
Big-Bang, esto es $cT$, donde $T\approx 10^{10} \mbox{ a\nn os} $ es
la edad del Universo. Vemos por lo tanto que la distancia al
horizonte, es decir la regi\oo n accesible a la observaci\oo n, aumenta
con el tiempo. Por lo tanto si, como todo parece indicar, la expansi\oo n
del Universo es eterna las regiones arbitrariamente alejadas de nosotros
acabar\a n siendo visibles en el futuro.

\section{Conclusiones}

En este breve ensayo hemos intentado transmitir las ideas b\a sicas
que constituyen la imagen relativista del mundo y como el trabajo de
Einstein revolucion\oo\ nuestra concepci\oo n del Universo en su
conjunto.  Con la relatividad especial aprendimos que el espacio y 
el tiempo absolutos de la f\ii sica newtoniana han de ser reemplazados 
por el espacio-tiempo, y que las leyes de la f\ii sica han de ser
compatibles con su estructura geom\e trica. Por
otra parte la Teor\ii a General de la Relatividad nos ha mostrado que
el espacio-tiempo no es un simple espectador d\oo nde los sucesos
tienen lugar sino que \e l mismo y su geometr\ii a son objetos de la
din\a mica.

Para el propio Einstein la formulaci\oo n de la relatividad general no
fue m\a s que el principio de la b\u squeda de una teor\ii a unificada
de la gravitaci\oo n y el electromagnetismo. En este empe\nn o su
principal herramienta fue de nuevo la geometr\ii a riemanniana y sus
generalizaciones. A pesar de que por diversas razones la empresa de
construir tal teor\ii a unificada estaba condenada al fracaso su
esp\ii ritu ha sobrevivido. Hoy en d\ii a una parte importante de la
comunidad cient\ii fica considera que la \u ltima frontera de la f\ii sica 
te\oo rica se encuentra precisamente en la consecuci\oo n de una descripci\oo n
unificada de todas las interacciones incluyendo la gravedad. Esto muy
probablemente implica la construcci\oo n de una teor\ii a cu\a ntica de
la gravedad cuyas propiedades hoy s\oo lo podemos atisbar.

Quiz\a s la mayor dificultad de esta s\ii ntesis final radique en que,
como hemos visto, la geometr\ii a del espacio-tiempo es juez y parte:
no s\oo lo nos proporciona el sustrato sobre el que la f\ii sica tiene
lugar, sino que al mismo tiempo participa en ella con su propia din\a
mica.  En la teor\ii a final el espacio-tiempo deber\a\ quiz\a s de
dejar de ser un concepto aprior\ii stico para convertirse en un
concepto emergente o derivado, igual que las part\ii culas lo son en
una teor\ii a cu\a ntica de campos. Una cumbre dificil de alcanzar,
sin duda, pero cuya culminaci\oo n se contar\a\ entre los m\a s grandes
logros de la mente humana.

\end{document}